\documentclass[useAMS,usenatbib,onecolumn]{mn2e}

\usepackage{dcolumn}
\usepackage{graphicx}
\usepackage{url}
\usepackage{color}
\usepackage{textcomp}
\usepackage[T1]{fontenc}

\newcommand{\cm}{cm$^{-1}$}

\newcommand{\eqref}[1]{(\ref{#1})}

\title[ExoMol molecular line lists X: NaH]{ExoMol molecular line lists X: The spectrum
of sodium hydride}

\date{\today}
\author[Rivlin et al]{Tom Rivlin$^{1}$, Lorenzo Lodi$^{1}$, Sergei N. Yurchenko$^{1}$,  Jonathan Tennyson$^{1}$,
\newauthor  Robert J. Le Roy$^{2}$ \\
$^{1}$Department of Physics and Astronomy, University College London, London WC1E 6BT, UK; \\
$^{2}$Department of Chemistry, University of Waterloo, Waterloo, Ontario N2L 3G1, Canada}
\date{Accepted XXXX. Received XXXX; in original form XXXX}

\pagerange{\pageref{firstpage}--\pageref{lastpage}} \pubyear{2014}

\begin{document}
\maketitle

\begin{abstract}

  Accurate and complete rotational, rotational-vibrational and
  rotational-vibrational-electronic line lists are calculated for sodium
  hydride: both the NaH and NaD isotopologues are considered. These line lists
  cover all ro-vibrational states of the ground ($X$~$^1\Sigma^+$) and first
  excited ($A$~$^1\Sigma^+$) electronic states. The calculations use available
spectroscopically-determined potential energy curves and new high-quality,
\textit{ab initio} dipole moment curves. Partition functions for both
isotopologues are
  calculated and the effect of quasibound states is considered.
 The resulting line lists are suitable for temperatures up to about 7000~K
  and are designed  for studies of exoplanet atmospheres, brown dwarfs and
  cool stars. In particular, the NaH $A-X$ band is found to show a broad
 absorption feature
  at about 385 nm which should provide a signature for the molecule.  All
partition functions, lines and transitions are
  available as Supplementary Information to this article and at \url{www.exomol.com}.

\end{abstract}

\textit{molecular data; opacity; astronomical data bases: miscellaneous; planets and satellites: atmospheres; stars: low-mass}

\label{firstpage}

\section{Introduction}

Although long discussed \citep{78KiDaxx.NaH}, the existence of sodium hydride, NaH, is yet to  be
confirmed in any astronomical bodies. Notably few metal hydrides have
been detected in the interstellar medium and searches for NaH in both dense clouds
\citep{82PlErxx.NaH} and diffuse clouds \citep{77SnSmxx.NaH,87CzFeRo.NaH}  have
so far only yielded upper limits. Similarly an attempt to detect the long-wavelength
signature for NaH in the atmosphere of  Jupiter and Saturn \citep{96WeSexx.NaH}
also proved negative.

While spectroscopic data for long-wavelength studies of NaH has been the
subject of the laboratory studies \citep{87LeZiEv.NaH,00OkTaxx.NaH}, and are
compiled in the CDMS database \citep{cdms}, the situation is less clear at
infrared and visible wavelengths. A recent work on M-dwarf models by
\citet{13RaReAl.NaHAlH} identified NaH as one of only three molecules for which
the necessary spectroscopic data is missing. This is despite the fact that the
$A-X$ band of NaH should give an observable feature at visible wavelengths.
It is this situation we address here. We note that the next lowest-lying
singlet electronic excited states of NaH are the $C$~$^1\Sigma^+$ \citep{15WaSeLe.NaH}
and the shallow  $B$~$^1\Pi$ state \citep{04YaZhHa.NaH} whose yet to be observed
spectra lie in the ultraviolet, probably at around 3000 \AA.

%\red{Importance of ultra-cold molecules?}
The large dipole moments of  NaH, as well
as supposedly making it more amenable to astronomical detection, has also
excited the interest of scientists working on ultra-cold molecules \citep{06JuPeKi.NaH,09AyDeDu.NaH,10SuYuNi.NaH}.
In particular, it has been
suggested that these dipoles should enhance the prospects of molecule formation by radiative association
at very low temperatures \citep{06JuPeKi.NaH}.

The ExoMol project aims to provide line lists of spectroscopic transitions for
key molecular species which are likely to be important in the atmospheres of
extrasolar planets and cool stars; its aims, scope and methodology have been
summarised by \citet{jt528}. In this paper, ro-vibrational and rovibronic
transition lists are computed for the only two stable isotopes of sodium
hydride, $^{23}$NaH and $^{23}$NaD.  These line lists consider
transitions within and between the two lowest electronic states of NaH,
$X$~$^1\Sigma^+$ and $A$~$^1\Sigma^+$. The resulting line lists are comprehensive
and should be valid up to the 7000~K temperature range.

\section{Method}

\begin{figure}
\begin{center}
\scalebox{0.4}{\includegraphics{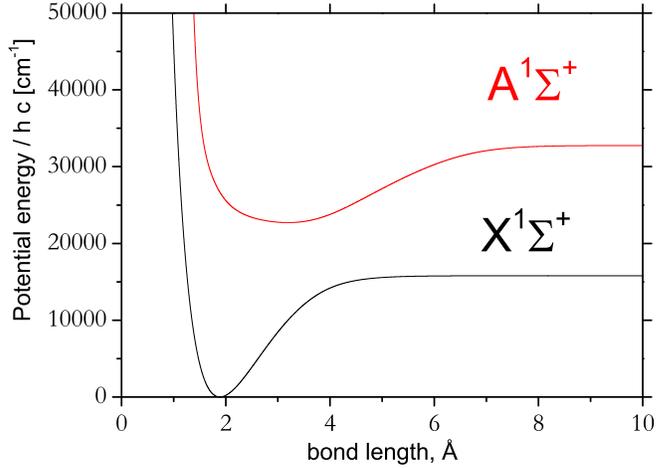}}
\caption{Empirical NaH $X ^{1}\Sigma^{+}_{g}$ and $A
^{1}\Sigma^{+}_{g}$ state potential energy curves due to \citet{15WaSeLe.NaH}.}
\label{fig:PECs}
\end{center}
\end{figure}

As with other diatomic systems studied as part of the ExoMol project
\citep{jt529,jt563,jt583,jt590,jt598}, we treated NaH using both \textit{ab initio}
and experimental data. \citet{15WaSeLe.NaH} used a direct potential fit
analysis to generate accurate potential energy curves for the ground and first
singlet excited states of NaH ($X$~$^{1}\Sigma^{+}_{g}$ and `Avoided Crossing' 
$A$~$^{1}\Sigma^{+}_{g}$). These curves (also shown in Fig.~\ref{fig:PECs}) were
then used as input to the program \textsc{LEVEL} version 8.2 \citep{lr07} to
produce accurate nuclear motion energy levels, wavefunctions and hence
transition intensities. A full discussion of how the potential energy curves
were derived
 can be found in \citet{15WaSeLe.NaH}, which also gives a comprehensive
 survey of previous laboratory work on the spectrum of NaH and NaD. We note that
the experimental data used by  \citet{15WaSeLe.NaH} to determine their curves
is insufficient to characterise the last 1000 \cm\ of the excited  $A$~$^{1}\Sigma^{+}_{g}$
below dissociation, where the A state undergoes another avoided crossing \citep{09AyDeDu.NaH}.
The dissocation limit of this state was constrained by  the atomic limit
neglecting spin-orbit coupling \citep{15WaSeLe.NaH} and ro-vibrational states computed
in this region must be considered approximate.

\subsection{Dipole Moments}

{\it Ab initio} dipole moments were computed on a dense, uniformly-spaced grid of
220 points from $r$=2.40 to $r$=13.40 a$_0$ in steps of 0.05 a$_0$.  Dipoles
were computed as expectation value of a multi-reference configuration
interaction (MRCI) calculation using a full valence complete
active space model, which distributes
the two outer electrons among 5 orbitals, and the  aug-cc-pwCV5Z basis set.
Orbitals were produced using a state-averaged complete active space self
consistent field (CASSCF) which considered both the $X$ and $A$ states. The
core 2s2p electrons were correlated at the MRCI stage. A single run takes about
4 minutes.

\begin{figure}
\begin{center}
\scalebox{0.4}{\includegraphics{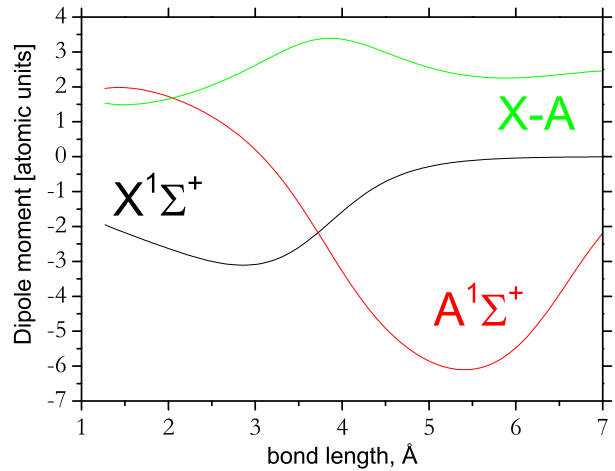}}
\caption{\textit{Ab initio} NaH dipole moment curves as a function of internuclear separation.}
\label{fig:dipoles}
\end{center}
\end{figure}

Figure~\ref{fig:dipoles} shows our two diagonal dipole moment curves and the
off-diagonal $A$--$X$ transition moment. The diagonal dipoles have a somewhat
unusual shape which is undoubted associated with the changing character of the
NaH wavefunction as a function of geometry. The dipole moment of ground state
NaH is has been determined experimentally \citep{79Daxxxx.NaH}, but not
particularly accurately. Conversely there have been quite a number of
theoretical studies on the problem. The most recent is a comprehensive
calculation of all three moment considered here by \citet{09AyDeDu.NaH}, who
also provide a comprehensive survey of the previous literature. Our results are
in excellent agreement with those \citet{09AyDeDu.NaH}, who also find the
various turning points in the dipole curves which are a feature of our
calculations.

Our dipole moment points, which are provided in the supplementary material,
were input directly into \textsc{LEVEL}.

\section{Results}

\subsection{Partition Function}

Partition functions for NaH and NaD were calculated by summing over all energy
levels calculated.  \citet{15SzCsxx.MgH} have recently shown that inclusion of
quasibound (resonance or pre-dissociative) states can influence the partition
function sums at higher temperatures. Here we present partition functions for
both NaH and NaD obtained by summing both over all the truly bound states of
the system and also by including quasibound states determined using
\textsc{LEVEL}. Table \ref{tab:Partition} compares these two methods and
suggests that below 2000 K the influence of quasibound states on the partition
sum can safely be neglected. Above this temperature the quasibound states have
a small but increasing influence. However for consistency with our previous
work and because we do not consider transitions to quasibound states, we ignore
the effect of quasibound states below. For high accuracy work at higher
temperatures this approximation may need to be re-visited.

 Table \ref{tab:Partition} also compares our
NaH partition functions with those of \citet{84SaTaxx.partfunc} and the CDMS
database \citep{cdms}. Our results cover a larger temperature range: CDMS values
only go up to 300~K for NaH, and Sauval \& Tatum only claim their values are
accurate for 1000 -- 9000~K. The latter values were multiplied by the factor 8
in order to explicitly account for the nuclear spin degeneracy in accord with
the HITRAN \citep{03FiGaGo.partfunc} convention, used by ExoMol; the CDMS
partition function had to be multiplied by 2. According with the
convention of CDMS, the common factors in partition functions are usually
divided to keep them small \citep{cdms}. After these adjustments results give
good agreement with Sauval \& Tatum and CDMS in the appropriate temperature
ranges, although there is a slight divergence from Sauval \& Tatum's values at
higher temperature ranges.

A polynomial fit of this work's partition functions is given in in Table
\ref{tab:Fitting}. The values in Table \ref{tab:Fitting} are coefficients in
the following equation:
\begin{equation}\label{eq:logexpansion}
\log_{10}Q(T) = \sum_{n=0}^{8}a_{n}(\log_{10}T)^n,
\end{equation}
which was based on a series expansion used by \citet{jt263}.

\begin{table}
\caption{Partition function of NaH compared with other sources}\label{tab:Partition}
\begin{tabular}{ c r r r r}
\hline\hline
  $T$(K) & This work & This work & \citet{84SaTaxx.partfunc}& CDMS \\
&(with quasibound) & (without quasibound)  \\
\hline
    37.5 &                  &                     &             51.00 &              45.93  \\
      38 &             46.50 &              46.50 &             51.67 &                     \\
      75 &             89.10 &              89.10 &             97.98 &              89.10  \\
     150 &            175.73 &             175.73 &            182.99 &             175.72  \\
     225 &            262.83 &             262.83 &            266.93 &             262.62  \\
     300 &            351.39 &             351.39 &            353.94 &             349.79  \\
     500 &            607.74 &             607.74 &            610.87 &                     \\
    1000 &           1479.85 &            1479.85 &           1460.52 &                     \\
    1500 &           2764.91 &            2764.89 &           2668.35 &                     \\
    2000 &           4521.88 &            4520.67 &           4303.28 &                     \\
    2500 &           6811.55 &            6799.15 &           6437.62 &                     \\
    3000 &           9692.63 &            9633.66 &           9149.42 &                     \\
    3500 &          13205.11 &           13024.25 &          12523.04 &                     \\
    4000 &          17363.28 &           16941.48 &          16649.49 &                     \\
    4500 &          22161.35 &           21341.29 &          21626.79 &                     \\
    5000 &          27583.33 &           26178.70 &          27560.23 &                     \\
    5500 &          33610.90 &           31415.33 &          34562.72 &                     \\
    6000 &          40227.09 &           37021.39 &          42755.06 &                     \\
  \hline
\end{tabular}
\end{table}

\begin{table}
\caption{Fitting parameters for the partition function of NaH/NaD.
Fits are valid for temperatures up to 9000~K within 1~\%.}
\label{tab:Fitting}
\begin{tabular}{ l r r r r}
\hline\hline
   & NaH & NaH & NaD & NaD\\
   &  (with quasibound) &  (without quasibound)& (with quasibound) &(without quasibound)\\
\hline
$a_0$    &     5.20312719799 &    -0.19230735479 &    -1.61166291473 &    -0.19230735479   \\
$a_1$    &   -19.01261024630 &     3.70859873334 &     9.53281763105 &     3.70859873334   \\
$a_2$    &    34.83255367250 &    -3.07887332529 &   -13.18395704240 &    -3.07887332529   \\
$a_3$    &   -34.37463416770 &     0.71923866264 &    10.40662225870 &     0.71923866264   \\
$a_4$    &    20.73051960960 &     1.11929540933 &    -4.49578453850 &     1.11929540933   \\
$a_5$    &    -7.76051252085 &    -0.99902883569 &     1.01674402973 &    -0.99902883569   \\
$a_6$    &     1.75196812228 &     0.34706733327 &    -0.09068752170 &     0.34706733327   \\
$a_7$    &    -0.21734537482 &    -0.05612884704 &    -0.00355980420 &    -0.05612884704   \\
$a_8$    &     0.01134093577 &     0.00348875707 &     0.00081789072 &     0.00348875707   \\
  \hline
\end{tabular}
\end{table}

\subsection{Line lists}

Line lists were calculated for the two main isotopologues: NaH and NaD. All
transitions satisfying the selection rule $\Delta J = \pm 1$ between all
ro-vibrational states in the electronic ground state and and first
electronically-excited state were considered.  A summary of each line list is
given in Table \ref{tab:Summary}. Although every possible transition was
computed, some were very weak and are not included in the final line list.

\begin{table}
\caption{Summary of the computed NaH and NaD line lists.}
\label{tab:Summary}
\begin{tabular}{ l r r r r}
\hline\noalign{\smallskip}
 & NaH $X$ state & NaH $A$ state & NaD $X$ state & NaD $A$ state\\\noalign{\smallskip}\hline
Maximum $v$ & 21 & 32 & 30 & 34\\
Maximum $J$ & 81 & 123 & 113 & 171\\
%Number of lines  & 1\,069 & 2\,380 & 2\,051 & 4\,563\\
  \hline
\end{tabular}
\end{table}

\begin{center}
\begin{table}
\caption{Extract from the states file for NaH.  The zero of energy is taken to be the energy of the lowest energy level.
The files contain  $3 \, 339$  levels for NaH and $5\, 960$ levels for NaD.}
\label{tab:StatesSample}
\begin{tabular}{ r r r r r r}
\hline\hline
$n$ & \multicolumn{1}{c}{$\tilde{E}$} &  $g$    & $J$      & State& $v$ \\
\hline
           1 &    0.000000  &    8  &   0.0 &  X    &       0  \\
           2 & 1133.102453  &    8  &   0.0 &  X    &       1  \\
           3 & 2228.214399  &    8  &   0.0 &  X    &       2  \\
           4 & 3285.975269  &    8  &   0.0 &  X    &       3  \\
           5 & 4306.940010  &    8  &   0.0 &  X    &       4  \\
           6 & 5291.562054  &    8  &   0.0 &  X    &       5  \\
           7 & 6240.146564  &    8  &   0.0 &  X    &       6  \\
           8 & 7152.840057  &    8  &   0.0 &  X    &       7  \\
           9 & 8029.619734  &    8  &   0.0 &  X    &       8  \\
          10 & 8870.226718  &    8  &   0.0 &  X    &       9  \\
          11 & 9674.057941  &    8  &   0.0 &  X    &      10  \\
          12 &10440.079797  &    8  &   0.0 &  X    &      11  \\
  \hline
\end{tabular}

$n$:   State counting number.\\
$\tilde{E}$: State energy in \cm.\\
$g$: State degeneracy.\\
$J$:   State rotational quantum number.\\
$v$:   State vibrational quantum number.\\
State: Electronic state label.\\

\end{table}
\end{center}

\begin{center}
\begin{table}
\caption{Extract from the .trans file for NaH. Full tables are available from the Exomol web site \protect\url{www.exomol.com} or from the
VizieR service from the web site of the Strasbourg  Data Center \protect\url{cds.u-strasbg.fr}.
The files contain $79~898$ transitions for NaH and $167~224$ transitions for NaD.}
\label{tab:TransSample}
\begin{tabular}{ r r r}
\hline\hline
$f$ & $i$ & $A_{if}$\\
\hline
     342    &    322   &   2.1771E+01 \\
    3190    &   3198   &   5.9304E+00 \\
    1168    &   1200   &   6.8873E-01 \\
    3329    &   3327   &   1.1725E+02 \\
     477    &    495   &   5.0419E-01 \\
    3225    &   3217   &   8.4415E+01 \\
    2887    &   2870   &   3.5561E+01 \\
     929    &    938   &   1.0760E-01 \\
     759    &    744   &   1.1033E+01 \\
    3286    &   3280   &   9.6810E+01 \\
  \hline
\end{tabular}

$f$: final state number.\\
 $i$: Initial state number.\\
$A_{if}$:  Einstein A coefficient in $s^{-1}$.\\

\end{table}
\end{center}

\section{Results}

The line lists contain hundreds of thousands of transitions. They are separated
into separate energy levels and transitions files. This is done using the
standard ExoMol format \citep{jt548} where the states file gives all energy
levels and associated quantum numbers, and  the transitions file gives Einstein
A coefficients and the numbers of the associated states. Extracts from the
start of the NaH states and transitions files are given in
 Tables \ref{tab:StatesSample} and \ref{tab:TransSample}.
The full line list can be
downloaded from the Exomol web site \url{www.exomol.com} or from the
VizieR service from the web site of the Strasbourg  Data
Center \url{cds.u-strasbg.fr}. A small computer program which uses
these files to compute spectra is given in the supplementary material
to \citet{jt590}.

The direct potential fit \citet{15WaSeLe.NaH} essentially reproduced all
experimental data within its stated uncertainty once a very modest amount of
data cleaning had been performed.
%Transition energies computed here are
%compared with some experimental measurements for transitions in the ground
%electronic state ($X$--$X$ transitions), the excited electronic state ($A$--$A$
%transitions), and for transitions from the excited to the ground state
%($A$--$X$ transitions).

%Table \ref{tab:ExptCompare} \red{(WILL BE UPDATED)} shows a comparison with
%sample measured frequencies for $X$-state pure rotational transitions
%\citep{87LeZiEv.NaH}, vibration-rotation transitions \citep{89MaOlxx.NaH} and
%electronic transitions \citep{97BaTsJi.NaH}. In all cases the agreement is
%excellent. Comparisons with $A$--$A$ experimental transition frequencies of
%\citet{49Paxxxx.NaH}, which are probably less important for astronomical
%studies, also gave good agreement. This suggests that, at least for transitions
%involving energy levels that have been probed experimentally, our transition
%frequencies should good to spectroscopic accuracy.

\citet{76Daxxxx.NaH}, \citet{76BaJoNa.NaH} and \citet{83NeGixx.NaH} measured
radiative life times for a number of $A$ state ro-vibrational levels. These
measurements suggest that the lifetime actually has little dependence on the
actual level involved with a sudden increase for $v=21$. A comparison with our
predictions for some of these states is listed in Table~\ref{tab:lifetime}: in
general the agreement is very good. However, in contrast to
\citet{83NeGixx.NaH}, we observe the increase of the life time of the $A$ state
levels at lower vibrational excitations than $v=21$. The increase is associated
with the $X$ state dissociation threshold of 15797.4~\cm, which truncates the
contributions from the bound states of the ground electronic state. Thus the
lower experimental values of the life times can be attributed to the
transitions to the $X$~$^1\Sigma^+$ quasibound states, which are not included in
our line lists.

%All our values show very little vibrational or rotational dependence.

%\citep{97LeHexx.NaH} determined relative dipole transition moments \red{useful? many not}

\begin{figure}
\begin{center}
\scalebox{0.4}{\includegraphics{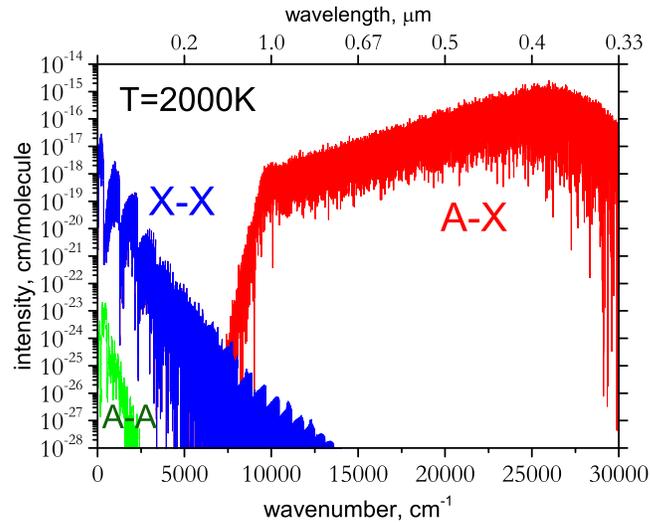}}
\caption{Overview of the $X$--$X$, $A$--$A$, and $A$--$X$ absorption spectrum of $^{23}$NaH at 2000~K.
The theoretical spectrum is convolved with  a Gaussian profile with a half-width half-maximum (HWHM) of 0.1~\cm.}
\label{fig:overview}
\end{center}
\end{figure}

\begin{figure}
\begin{center}
\scalebox{0.4}{\includegraphics{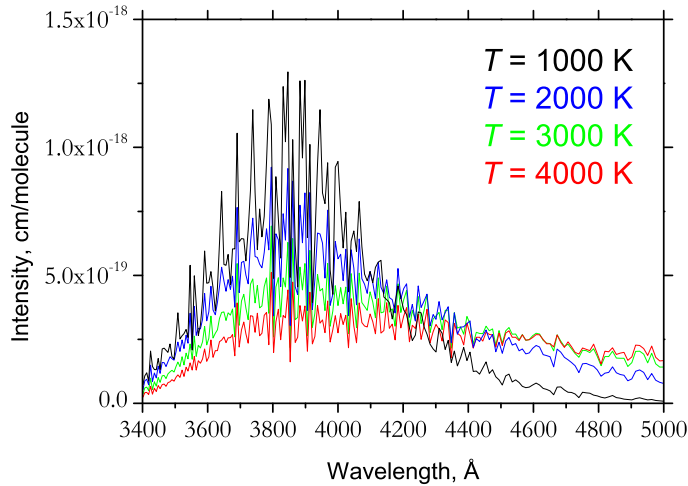}}
\caption{Computed $A$--$X$ absorption spectrum of $^{23}$NaH at different temperatures.
A Doppler profile was used to convolve the
theoretical spectrum. Wavenumber bin of 50~\cm\ was chosen to mimic resolving power of about  500 to 1000.}
\label{fig:AX}
\end{center}
\end{figure}

Figure  \ref{fig:overview} gives an overview of the entire spectral range
covering the $X$ and $A$  systems, from infrared to visible shown as an
absorption spectrum generated for $T=2000$~K.  Of particular astronomical
interest is likely to be the $A - X$ band in the 380~nm region.
 The $A - X$ band does not exhibit sharp features which can be attributed to the shallow $A$ potential energy curve, see
Fig.~\ref{fig:PECs}. This reduces the chances of NaH to be detected at visible wavelengths,
especially at high temperature, as also illustrated by Fig.~\ref{fig:AX}, where
the $A - X$ absorption band is shown for a set of temperatures from $1000$ to
$4000$~K.

\begin{figure}
\begin{center}
\scalebox{0.4}{\includegraphics{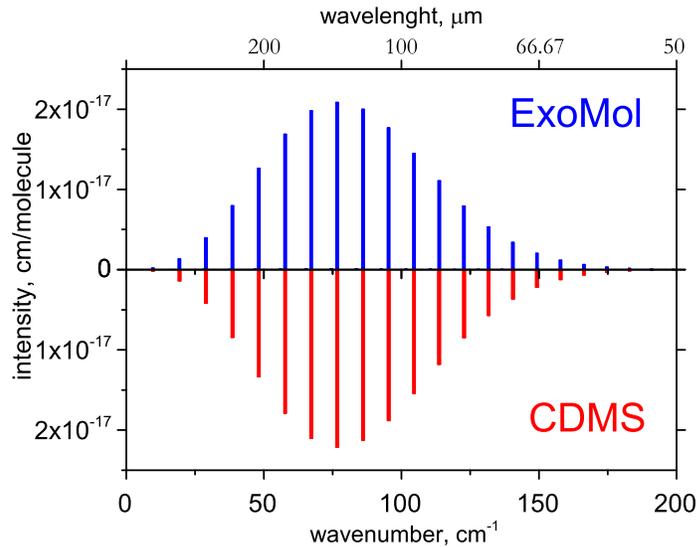}}
\caption{Rotational spectrum of $^{23}$NaH computed at $T=298$~K: ExoMol (current work) and CDMS \citep{cdms}.
The CDMS intensities were obtained by summing  individual hyperfine transitions; these
were generated by \citet{cdms} using an equilibrium dipole moment of 6.7~D as calculated by
\citet{75SaHiSa.NaH}.}
\label{fig:CDMS}
\end{center}
\end{figure}

Figure~\ref{fig:CDMS} shows the pure rotational band of NaH at $T=298$~K
compared to the spectra from the CDMS database. The latter was generated by
averaging over the the hyperfine lines, which are not resolved in our line
lists. The agreement between intensities can be attributed to the  equilibrium
$X$-state dipole moment value of 6.7~D of \citet{75SaHiSa.NaH} used by CDMS. Our
value is 6.43~D at $r_{\rm e}=1.887$.

\begin{table}\footnotesize
\caption{Lifetimes (in ns) of selected ro-vibrational states within the $A$~$^1\Sigma^+$ electronic state: comparison
of experiment \citet{76Daxxxx.NaH,83NeGixx.NaH} and theory (this work).}
\label{tab:lifetime}
\begin{tabular}{rrcr@{}lr}
\hline\hline
$v^{\prime}$     &  $J^{\prime}$       &   Ref.  &    \multicolumn{2}{c}{ Exp.}   &   Calc.           \\
\hline
    3 &        8  &   \citet{76Daxxxx.NaH}        &24.0     &   (3)    &       25.71  \\%      25.50
    4 &       11  &   \citet{76Daxxxx.NaH}        &28.3     &   (3)    &       25.69  \\%      25.47
    5 &       16  &   \citet{76Daxxxx.NaH}        &27.1     &   (3)    &       26.44  \\%      25.98
    3 &        4  &   \citet{83NeGixx.NaH}        &27     &          &       25.37  \\%      25.16
    4 &        6  &   \citet{83NeGixx.NaH}        &     28.5    &          &       25.16  \\%      24.96
    7 &        6  &   \citet{83NeGixx.NaH}        &     26.5    &          &       25.40  \\%      26.22
    9 &        6  &   \citet{83NeGixx.NaH}        &      26     &          &       29.71  \\%      24.88
   11 &        6  &   \citet{83NeGixx.NaH}        &     25.5    &          &       33.10  \\%      25.14
   14 &        7  &   \citet{83NeGixx.NaH}        &      27     &          &       36.49  \\%      25.59
   18 &       11  &   \citet{83NeGixx.NaH}        &     27.5    &          &       44.01  \\%      26.06
   21 &        6  &   \citet{83NeGixx.NaH}        &      35     &          &       48.67  \\%      25.78
\hline
\end{tabular}
\end{table}

\section{Conclusions}

We present comprehensive line lists for
 NaH and NaD. These are based on the direct solution of the
nuclear motion Schr\"{o}dinger equation using potential energy curves obtained
by fitting to experimental data of measured transitions, all calculated by
\citet{15WaSeLe.NaH}. These were combined with a new \textit{ab initio} dipole
moment curve to produce comprehensive line lists for these two species. The
strong $A$--$X$ feature about 380 nm should provide a  signature for NaH in the
atmospheres of cool stars, brown dwarfs and exoplanets. However, the shallow
nature of the upper $A$~$^1\Sigma^+$ electronic state meansi that this feature
is predicted to be rather broad. While this means that NaH can make a
significant contribution to the opacity, it is likely to make this feature
difficult to detect.

NaH was recently identified one of only two key diatomic species
whose spectral data was completely missing from M-dwarf models \citep{13RaReAl.NaHAlH}.
A study on the other of these species, AlH, has just been completed and will be reported soon.

\section*{Acknowledgements}

 This work was supported  by the ERC under
the Advanced Investigator Project 267219 and a Wolfson Royal Society
Research Merit award.

\bibliographystyle{mn2e}
%\bibliography{journals_astro,jtj,NaH,MgH,linelists,methods,additional,abinitio}

%\end{thebibliography}

\label{lastpage}

\end{document}